\documentclass{paper}

\preprintnumber{TTP12-024}

\DeclareFontShape{OT1}{cmtt}{bx}{n}{<5><6><7><8><9><10><10.95><12><14.4><17.28><20.74><24.88>cmttb10}{}

\newcommand{\myFitter}{\textit{my}Fitter}
\newcommand{\code}[1]{\texttt{#1}}
\newcommand{\Lcal}{\mathcal{L}}
\DeclareMathOperator{\CDF}{CDF}
\DeclareMathOperator{\Span}{span}
\DeclareMathOperator{\Erf}{Erf}

\title{Numerical Computation of $\boldsymbol p$-values with \myFitter}

\author{%
  M.\ Wiebusch%
  \email{wiebusch@particle.uni-karlsruhe.de}%
}

\institute{%
  Institute for Theoretical Particle Physics,\\
  Karlsruhe Institute of Technology (KIT), D-76128 Karlsruhe, Germany}

\abstract{Likelihood ratio tests are a widely used method in global analyses in
  particle physics. The computation of the statistical significance ($p$-value)
  of these tests is usually done with a simple formula that relies on Wilks'
  theorem. There are, however, many realistic situations where Wilks' theorem
  does not apply.  In particular, no simple formula exists for the comparison of
  models that are not \emph{nested}, in the sense that one model can be obtained
  from the other by fixing some of its parameters. In this paper I present
  methods for efficient \emph{numerical} computations of $p$-values, which work
  for both nested and non-nested models and do not rely on additional
  approximations. These methods have been implemented in a publicly available
  C++ framework for maximum likelihood fits called {\myFitter} and have recently
  been applied in a global analysis of the Standard Model with a fourth
  generation of fermions.}

\begin{document}
\maketitlepage

\section{Introduction}
\label{sec:intro}

Even though the LHC experiments have, so far, not found any clear signs for
physics beyond the Standard Model (SM) they already put strong constraints on
the favourite SM extensions of many theorists. For example, the SM with a
(perturbative) sequential fourth generation of fermions (SM4) has recently been
excluded at the $5\sigma$ level by a combination of Higgs and electroweak
precision data \cite{Eberhardt:2012gv} (see also \cite{Eberhardt:2012sb,
  Eberhardt:2012ck}). Other models with additional fermions or even some
constrained versions of Supersymmetry may follow soon.

In this situation some thoughts should be spent on the methods and criteria by
which we decide if a certain model is ruled out. A well-established technique in
(frequentist) statistical analyses is the method of \emph{likelihood ratio
  tests}. (For an introduction see e.g.\ \cite{CasellaBerger} or the statistics
chapter of \cite{PDG}.) In this method two models are compared with a test
statistic constructed from the ratio of their likelihood functions.  Wilks'
theorem states that under certain assumptions the test statistic is distributed
according to the well-known $\chi^2$-distribution \cite{Wilks}. In this case the
relation between the likelihood values at the best fit points and the
statistical significance ($p$-value) of the corresponding hypothesis test is
described by the normalised lower incomplete gamma function.

There are, however, also many realistic scenarios where Wilks' theorem does not
hold and the probability density function of the test statistic is not known
analytically. One example is the case of likelihood ratio tests where the two
models to be compared are not \emph{nested}, meaning that one model can not be
obtained from the other by fixing some of its parameters.  This problem was
encountered in the above-mentioned analyses of the Standard Model (SM) with a
fourth generation of fermions \cite{Eberhardt:2012gv, Eberhardt:2012sb,
  Eberhardt:2012ck}.  In these analyses it is not possible to regard the SM with
three fermion generations as a limiting case of the SM with four generations due
to non-decoupling contributions of chiral fermions in electroweak precision
observables and Higgs production and decay rates. Another case where analytical
formulae for $p$-values are not reliable is the situation where some of the
parameters of a model are bounded, in the sense that they are only allowed to
float within a certain range. Most notably, this applies to analyses where
systematic errors are treated within the $R$Fit scheme \cite{Hocker:2001xe},
i.e.\ by introducing so-called nuisance parameters with a limited range.

When analytic formulae fail one has to resort to numerical methods, and the
computation of $p$-values is no exception. The brute-force method is to generate
a large sample of random \emph{toy measurements} distributed according to the
prediction of the null hypothesis. For each toy measurement the value of the
test statistic is computed and compared to the value obtained from the actual
data. With a large enough sample we can then estimate the probability that the
value of the test statistic is larger than a certain number, usually chosen to
be the value of the test statistic obtained from the observed data. This
probability is called \emph{statistical significance} or \emph{$p$-value} of the
test. Unfortunately, the computational cost of the required numerical
simulations can be rather high, especially when the $p$-value is small. In this
paper I discuss some methods for improving the efficiency of numerical
computations of $p$-values. These methods have been applied in
\cite{Eberhardt:2012ck, Eberhardt:2012gv}, where, based on the constraints from
Higgs searches and electroweak precision observables, likelihood ratio tests
comparing the SM with three and four fermion generations were performed. The
methods are also implemented in a publicly available code called {\myFitter},
which I present in this paper.

The paper is organised as follows: in Sec.~\ref{sec:setup} I describe the
general mathematical setup and the definitions of the test statistics for nested
and non-nested models.  In Sec.~\ref{sec:ana} I sketch the derivation of Wilks'
theorem and discuss its range of applicability. In Sec.~\ref{sec:num} I explain
the strategy for improving the efficiency of numerical computations of
$p$-values. The {\myFitter} framework and the implementation of the methods from
Sec.~\ref{sec:num} are discussed in Sec.~\ref{sec:myFitter}.  Performance tests
of the {\myFitter} code are presented in Sec.~\ref{sec:performance}. I conclude
in Sec.~\ref{sec:concl}.

\section{General Setup}
\label{sec:setup}

Let $\vec X=(X_1,\ldots,X_n)$ be a set of experimental observables. In
frequentist statistics we regard observables as random variables distributed
according to some probability density function (PDF). A statistical model with
free parameters $\vec\xi=(\xi_1,\ldots,\xi_k)$ is therefore described by a
function $f(\vec x,\vec\xi)$, which must be a PDF for any fixed value of
$\vec\xi$ and considered as a function of $\vec x$ only:
\begin{equation}\label{eq:fnorm}
  \int d^n\vec x\, f(\vec x,\vec\xi) = 1
  \eqsep.
\end{equation}
The problem of statistical inference is to draw conclusions about the parameters
$\vec\xi$ from a given set of measurements $\vec x$ of the observables $\vec X$.

In global analyses like \cite{Eberhardt:2012gv, Eberhardt:2012sb,
  Eberhardt:2012ck} the observables come from many different collider
experiments and the parameters $\vec\xi$ to be determined are the fundamental
parameters of some theory of particle physics (the SM or extensions thereof). In
this situation the function $f$ usually does not depend on the parameters
$\vec\xi$ directly.  For example, a cross section $\sigma$ is measured by
counting the number $N$ of events that pass certain cuts and dividing by the
integrated luminosity $\Lcal$ and the selection efficiency $\epsilon$. If the
theory is realised with parameters $\vec\xi$ we denote the predicted value of
the cross section as $\tilde\sigma(\vec\xi)$. The integrated luminosity and
selection efficiency are usually constants which, to a good approximation, do
not depend on the theory parameters. Since $N$ follows a Poisson distribution
with mean value $\Lcal\epsilon\tilde\sigma(\vec\xi)$ the distribution of the
\emph{measured} value of the cross section, $\sigma=N/(\Lcal\epsilon)$, depends
on $\vec\xi$ only trough the \emph{predicted} cross section
$\tilde\sigma(\vec\xi)$.

These considerations motivate us to write the function $f$ in the following way:
\begin{equation}\label{eq:f}
  f(\vec x,\vec\xi) = \exp[-\tfrac12D(\tilde{\vec x}(\vec\xi),\vec x)]
  \eqsep,
\end{equation}
where $\tilde{\vec x}(\vec\xi)$ are the ``predicted'' values of the observables
and $D$ is a function which we shall call the \emph{input function}. The goal of
this re-writing is to cleanly separate information about the theoretical model
from information about the experimental uncertainties: the theory is described
by the function $\tilde{\vec x}$, which maps parameters to observables, and the
experimental uncertainties are described by the function $D(\tilde{\vec x},\vec
x)$, which is $-2$ times the logarithm of the probability density for measuring
values $\vec x$ if the ``true'' values are $\tilde{\vec x}$. For example, in
many situations the experimental errors are Gaussian, independent of the true
values $\tilde{\vec x}$ and described by a covariance matrix $V$. In this case
we have
\begin{equation}\label{eq:gaussian}
    D(\tilde{\vec x},\vec x)
  = (\tilde{\vec x}-\vec x)^\trans V^{-1}(\tilde{\vec x}-\vec x) + n\ln(2\pi)
  \eqsep,
\end{equation}
which, if substituted in \eqref{eq:f}, gives the usual expression for a
correlated Gaussian probability density with central value $\tilde{\vec
  x}(\vec\xi)$. Up to a constant term, $D(\tilde{\vec x}(\vec\xi),\vec x)$ is
simply the $\chi^2$-value associated with the parameters $\vec\xi$, input data
$\vec x$ and covariance matrix $V$.

For the methods presented in this paper, the exact definition of the functions
$\tilde{\vec x}$ and $D$ is not important, as long as $D$ has the following
properties:
\begin{enumerate}
\item For any given $\tilde{\vec x}$ the input function $D$ must
  satisfy the normalisation condition
  \begin{equation}
    \int_S d^n\vec x\,e^{-\frac12D(\tilde{\vec x},\vec x)} = 1
    \eqsep.
  \end{equation}
\item For any given $\vec x$, and considered as a function of $\tilde{\vec x}$,
  the input function $D(\tilde{\vec x},\vec x)$ must be bounded from below and
  have its unique absolute minimum at $\tilde{\vec x}=\vec x$.
\item For any given $\tilde{\vec x}$, and considered as a function of $\vec x$,
  the input function $D(\tilde{\vec x},\vec x)$ must be bounded from below
  and have its unique absolute minimum at $\vec x=\tilde{\vec x}$.
\end{enumerate}
The first property guarantees that \eqref{eq:fnorm} is satisfied.
The second property guarantees that, if parameters $\hat{\vec\xi}$ exist
with $\tilde{\vec x}(\hat{\vec\xi})=\vec x$, the maximum likelihood estimate
of the parameters is indeed $\hat{\vec\xi}$.
The third property can be regarded as a definition of the term ``predicted
value'': if the theory is realised with some parameters $\vec\xi$, the most
likely outcome of a measurement of the observables $\vec x$ should be $\vec
x=\tilde{\vec x}(\vec\xi)$.

Note that, without any modifications to the model, the third property does
\emph{not} hold in the presence of systematic errors, since a systematic error
is an offset between the true value of an observable and its most likely
measured value. This offset is the same each time the measurement is performed
and does therefore not average out when the measurement is repeated many
times. This results in a difference between $\vec x$ and the maximum of the
distribution of the random variables $\vec X$. The central idea of the $R$Fit
method \cite{Hocker:2001xe} is that systematic errors should not be treated as
errors at all, but as unknown theory parameters, so-called \emph{nuisance
  parameters}, that may vary within a certain range. In a way, the presence of a
systematic error means that theorists and experimentalists are simply not
talking about the same quantity.  Since the difference between the two
quantities can neither be modeled nor measured it has to be treated as an
additional model parameter, but with a limited range of possible values. Thus,
the third assumption \emph{does} hold if systematic errors are treated within
the $R$Fit scheme, i.e. by introducing a nuisance parameter for each source of
systematic errors.

The results of hypothesis tests for a certain theoretical model should not
depend on the way we parametrise the model. To make this
parametrisation-independence manifest it is convenient to define the
\emph{theory manifold} as the image of the function $\tilde{\vec x}$:
\begin{equation}
  M = \{\tilde{\vec x}(\vec\xi)\,|\,\vec\xi\in\Omega\}
  \eqsep,
\end{equation}
where $\Omega\subset\R^k$ is the \emph{parameter space} (i.e. the set of allowed
parameter values) of the model. Different parametrisations of the same model
are represented by different functions $\tilde{\vec x}$ and parameter
spaces $\Omega$, but always have the same theory manifold.

The general procedure for a likelihood ratio test (LRT) with \emph{nested}
models may now be described as follows: given certain experimental data $\vec
x$, we first maximise the PDF $f(\vec x,\vec\xi)$ with respect to the parameters
$\vec\xi$. This is equivalent to minimising the function $D(\tilde{\vec x},\vec
x)$ with respect to $\tilde{\vec x}$ on the theory manifold $M$:
\begin{equation}\label{eq:Dmin}
  f^\text{max}(\vec x) = \exp[-\tfrac12 D^\text{min}(\vec x)]
  \eqsep\text{with}\eqsep
  D^\text{min}(\vec x)
  = \min\{D(\tilde{\vec x},\vec x)\,|\,\tilde{\vec x}\in M\}
  \eqsep.
\end{equation}
Next, we consider a \emph{constrained} version of the model, which is usually
obtained from the original model by fixing some of its parameters. However, with
the notion of theory manifolds at hand, we can be more general and simply
require that the theory manifold $M_c$ of the constrained model is a subset of
$M$:
\begin{equation}
  M_c\subset M
  \eqsep.
\end{equation}
Maximising the likelihood for the constrained model we get
\begin{equation}\label{eq:Dcmin}
  f_c^\text{max}(\vec x) = \exp[-\tfrac12 D_c^\text{min}(\vec x)]
  \eqsep\text{with}\eqsep
  D_c^\text{min}(\vec x)
  = \min\{D(\tilde{\vec x},\vec x)\,|\,\tilde{\vec x}\in M_c\}
  \eqsep.
\end{equation}
Now we construct a test statistic $S$ from the ratio of the two maximum
likelihood values:
\begin{equation}\label{eq:S}
  S(\vec x) = -2\ln\frac{f_c^\text{max}(\vec x)}{f^\text{max}(\vec x)}
            = D_c^\text{min}(\vec x) - D^\text{min}(\vec x)
  \eqsep.
\end{equation}
To perform the actual test, we choose a certain realisation of the constrained
model as null hypothesis. Let $\vec\xi_0$ be the corresponding parameters and
$\tilde{\vec x}_0=\tilde{\vec x}(\vec\xi_0)$. The
statistical significance, or $p$-value, of the test is obtained by considering
an ensemble of \emph{toy measurements} $\vec x$ distributed according
to the PDF $f(\vec x,\vec\xi_0)$ and computing the probability that
$S(\vec x)$ is larger than some threshold value $S_0$:
\begin{equation}\label{eq:pvalue}
  p = \int d^n\vec x\,f(\vec x,\vec\xi_0)\theta(S(\vec x)-S_0)
  \eqsep,
\end{equation}
where $\theta$ denotes the Heavyside step-function. If, in the real experiments,
the data $\vec x_0$ was measured, one typically takes the maximum likelihood
estimates for $\vec x_0$ in the constrained model as null hypothesis, i.e. one
chooses $\vec\xi_0$ so that $\tilde{\vec x}_0\in M_c$ and $D(\tilde{\vec
  x}_0,\vec x_0)=D_c^\text{min}(\vec x_0)$. Then one performs the test with
$S_0=S(\vec x_0)$.

Note that the definition \eqref{eq:pvalue} is manifestly independent of the
parametrisation of the models: the factor $f(\vec x,\vec\xi_0)$ is fixed by the
null hypothesis and the test statistic is defined in terms of the functions
$D^\text{min}$ and $D_c^\text{min}$ whose definitions \eqref{eq:Dmin} and
\eqref{eq:Dcmin} depend on the manifolds $M$ and $M_c$, but not on their
parametrisation. This parametrisation-independent language allows us to easily
generalise the definition \eqref{eq:pvalue} for the case of (a large class of)
non-nested models.  Consider two models with theory manifolds $M_1$ and $M_2$
such that $M_1\not\subset M_2$ and $M_2\not\subset M_1$. However, we assume that
the relation between the PDFs of the two models and their respective theory
manifolds is still given by \eqref{eq:f} with the \emph{same} input function
$D$. This usually holds for global fits in particle physics, where a model
imposes certain relations between the predicted observables, but the random
distribution of the measured quantities is fixed by the predicted values,
irrespective of the model under consideration. In this case we can simply
combine the two theories into one by joining their theory manifolds,
\begin{equation}
  M \equiv M_1\cup M_2
  \eqsep\Rightarrow\eqsep
  M_1,M_2\subset M
  \eqsep,
\end{equation}
and do a LRT as described above, with $M$ as the full theory and either $M_1$
or $M_2$ as the constrained theory. Let
\begin{equation}
   D_1^\text{min}(\vec x)
  =\min\{D(\tilde{\vec x},\vec x)\,|\,\tilde{\vec x}\in M_1\}
  \eqsep,\eqsep
   D_2^\text{min}(\vec x)
  =\min\{D(\tilde{\vec x},\vec x)\,|\,\tilde{\vec x}\in M_2\}
  \eqsep.
\end{equation}
Then the test statistic for testing $M_2$ against $M$ is
\begin{equation}\label{eq:S2}
  S_2(\vec x) = \begin{cases}
    D_2^\text{min}(\vec x)-D_1^\text{min}(\vec x)
      &\text{for}\eqsep D_1^\text{min}(\vec x) < D_2^\text{min}(\vec x) \\
    0 &\text{otherwise}
  \end{cases}
\end{equation}
and the test statistic for testing $M_1$ against $M$ is obtained by exchanging
$D_1^\text{min}$ and $D_2^\text{min}$. Assume without restriction that for the
measured data $\vec x_0$ we have
\begin{equation}
  D_1^\text{min}(\vec x_0)\leq D_2^\text{min}(\vec x_0)
  \eqsep.
\end{equation}
Then $S_1(\vec x_0)$ is zero and the LRT for $M_1$ (using $S_1(\vec x_0)$ as
threshold value for the test) has a $p$-value of 1.  So, only the LRT for the
model which describes the data less well (i.e.\ gives a bigger value for
$D_i^\text{min}(\vec x_0)$) can have a $p$-value smaller than one.

\section{Analytical Formulae for \texorpdfstring{$\boldsymbol p$}{p}-values}
\label{sec:ana}

In many cases the computation of $p$-values in LRTs is trivial due to a theorem
by Wilks \cite{Wilks}. It states that the test statistic $S$ from \eqref{eq:S}
follows a $\chi^2$ distribution with $\dim(M)-\dim(M_c)$ degrees of freedom
\emph{if} the models are nested and the maximum likelihood estimates
$\hat{\vec\xi}(\vec x)$ of the parameters $\vec\xi$ follow a Gaussian
distribution. In this case the $p$-value \eqref{eq:pvalue} is given by
\begin{equation}\label{eq:pvalue-Wilks}
  p = 1 - P_{\nu/2}(S_0/2)
  \eqsep,
\end{equation}
where $\nu=\dim(M)-\dim(M_c)$ is the difference of dimensions of the theory
manifolds (usually equal to the number of parameters that were fixed) and
$P$ denotes the normalised lower incomplete Gamma function.

In global analyses in particle physics Wilks' theorem is commonly used, but the
validity of its underlying assumptions are rarely discussed. For PDFs of the
form \eqref{eq:f} the requirements for Wilks' theorem translate to certain
assumptions about the function $D$ and the theory manifolds $M$ and
$M_c$. These assumptions are:
\begin{description}
\item[Gaussianity.] The function $D(\tilde{\vec x},\vec x)$ only depends on the
  difference $\tilde{\vec x}-\vec x$ and is quadratic in this difference.
\item[Linearity.] The theory manifolds $M$ and $M_c$ are \emph{hyperplanes}.
\item[Nestedness.] The constrained theory is a subset of the full theory:
  $M_c\subset M$.
\end{description}
The first assumption, combined with the properties of $D$ discussed in
Sec.~\ref{sec:setup}, implies that $D$ is of the form \eqref{eq:gaussian},
i.e. that the experimental errors are Gaussian. The second assumption is invalid
if experimental errors are large, so that the curvature of the theory manifolds
can not be neglected. It also fails if some parameters of the model have upper
or lower bounds, so that the corresponding manifold does not extend to
infinity. The last assumption is invalid if none of the two models to be
compared can be considered as a special case of the other model.

\begin{figure}
  \centering\includegraphics{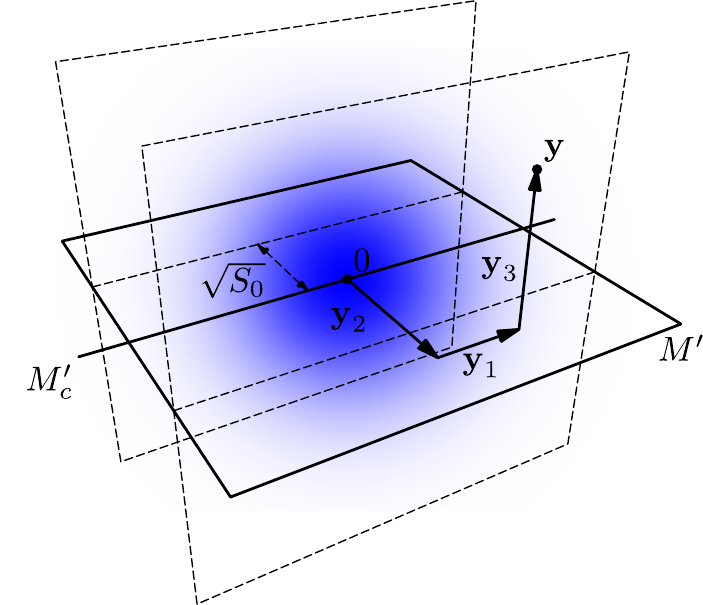}
  \caption{Derivation of Wilks' theorem. The blue colour indicates the
    probability density in the transformed space of observables $\vec y$.  It is
    an $n$-dimensional normal distribution. The $\theta$ function in
    \eqref{eq:pvalue} vanishes in the region between the planes indicated by the
    thin dashed lines.}
  \label{fig:nested}
\end{figure}
The derivation of Wilks' theorem from the assumptions above will be instructive
for our discussion of numerical methods in the next section, so I will briefly
sketch it here. The first step is to perform an affine-linear coordinate
transformation in the space of observables, which maps $\tilde{\vec x}_0$
(the predicted observables under the null hypothesis) to the origin
and changes the PDF to an $n$-dimensional normal distribution. In other words,
we introduce new coordinates $\vec y\equiv\vec y(\vec x)$, so that
$\vec y(\tilde{\vec x}_0)=0$ and
\begin{equation}
  f(\vec x,\vec\xi_0)=\frac1{(2\pi)^{n/2}}e^{-\|\vec y\|^2/2}
  \eqsep\Rightarrow\eqsep
  D(\tilde{\vec x}_0,\vec x) = \|\vec y\|^2 + n\ln(2\pi)
  \eqsep,
\end{equation}
The linear part of this transformation is easily constructed by diagonalising
the matrix $V$ from \eqref{eq:gaussian} and then scaling the new coordinates
appropriately.  We see that, up to a constant term, the function $D$ is simply
the squared euclidean norm of the vector $\vec y$ (denoted as $\|\vec
y\|^2$). Let $M'$ and $M_c'$ denote the images of $M$ and $M_c$ under the
coordinate transformation $\vec y$. Since $M$ and $M_c$ both contain
$\tilde{\vec x}_0$, the hyperplanes $M'$ and $M_c'$ both contain the origin and
are therefore linear subspaces. Consequently, the functions $D^\text{min}$ and
$D_c^\text{min}$ (see Eq.~\ref{eq:Dmin} and \ref{eq:Dcmin}) are simply the
squared euclidean length of the component of $\vec y$ perpendicular to $M'$ and
$M_c'$, respectively. For $n=3$, $\dim(M)=2$ and $\dim(M_c)=1$ the situation
is depicted in Fig.~\ref{fig:nested}. We may write the vector $\vec y$
as a sum of three orthogonal vectors $\vec y_1$, $\vec y_2$ and $\vec y_3$
with $\vec y_1\in M_c'$ and $\vec y_2\in M$. The test statistic $S$
is then:
\begin{equation}
  S(\vec x) = \|\vec y_2+\vec y_3\|^2 - \|\vec y_3\|^2
            = \|\vec y_2\|^2
  \eqsep.
\end{equation}
In terms of the coordinates $\vec y$ the integral from \eqref{eq:pvalue}
becomes
\begin{equation}\label{eq:pvalue-Wilks-1}
  p = \frac1{(2\pi)^{n/2}}\int d^n\vec y\, e^{-\|\vec y\|^2/2}
      \theta(\|\vec y_2\|^2-S_0)
\end{equation}
In other words, the $p$-value is the integral of an $n$-dimensional normal
distribution in the region outside an (infinitely long) ``hyper-cylinder''
defined by $\|\vec y_2\|^2>S_0$. In Fig.~\ref{fig:nested}, this
``cylinder'' is the region between the planes indicated by dashed lines.  The
integral in \eqref{eq:pvalue-Wilks-1} can easily be computed. The integrals over
the components $\vec y_1$ and $\vec y_3$ are just Gaussian integrals and give an
overall factor of $(2\pi)^{(n-\nu)/2}$. Introducing spherical coordinates in the
$\nu$-dimensional subspace corresponding to the component $\vec y_2$ and
exploiting rotational symmetry immediately leads to \eqref{eq:pvalue-Wilks}.

\section{Numerical Calculation of \texorpdfstring{$\boldsymbol p$}{p}-values}
\label{sec:num}

In the last section we have seen how Wilks' theorem emerges from geometric
arguments if the models under consideration satisfy three assumptions, which we
called \emph{gaussianity}, \emph{linearity} and \emph{nestedness}. In practice,
these assumptions are rarely satisfied exactly. Usually, they are only more or
less valid approximations. If we do not want to rely on these approximations we
have to resort to numerical integration methods to calculate $p$-values. To make
these methods efficient it is a good idea to take some or all of the
approximations as a starting point and optimise the numerical integration for
the case where they are valid.  In the following, we will use Monte Carlo
integration with importance sampling to compute the integral \eqref{eq:pvalue},
and construct sampling densities which are optimal for models that satisfy
gaussianity and linearity.

To compute the integral \eqref{eq:pvalue} with the importance sampling method,
we generate a large number $N$ of sample points $\vec x_i$ according to some
sampling distribution $\rho$. The integral \eqref{eq:pvalue} is then estimated
as
\begin{equation}\label{eq:pvalueMC}
  p = \frac1N\sum_{i=1}^N\frac{f(\vec x_i,\vec\xi_0)}{\rho(\vec x_i)}
                         \theta(S(\vec x_i)-S_0)
  \eqsep.
\end{equation}
To reduce the statistical error of this estimate one has to choose the function
$\rho$ as similar as possible to the integrand, so that the terms in the sum are
(ideally) all of the same size. A common approach (especially if
$f(\cdot,\vec\xi_0)$ is a Gaussian distribution) is to choose $\rho(\vec
x)\approx f(\vec x,\vec\xi_0)$ and let the numerics take care of the theta
function in the integrand. For large $p$-values this is a viable option, but if
$p$ is small most sample points give a contribution of zero to the integrand and
the numerical integration becomes very inefficient. (Remember that each
evaluation of the test statistic $S(\vec x)$ requires the computation of
$D^\text{min}(\vec x)$ and $D_c^\text{min}(\vec x)$ which, in general, has to be
done by numerical minimisation.) For small $p$-values, the efficiency of the
integration can be significantly improved by choosing a sampling density $\rho$
which avoids the region where the theta function is zero (i.e.\ the region
between the dashed planes in Fig.~\ref{fig:nested}). Knowledge about the
geometric properties of the theory manifolds can be used to construct such a
sampling density. In the numerical methods proposed in this paper, we assume
gaussianity and linearity for the purpose of constructing the sampling density,
but make no approximations when computing the $p$-value.

To see how this works, let us start with the case where the nestedness
assumption is still valid. For definiteness, we choose a parametrisation
so that
\begin{equation}
  M=\{\tilde{\vec x}(\vec\xi)\,|\,\vec\xi\in\Omega\}
  \eqsep,\eqsep
  M_c=\{\tilde{\vec x}(\vec\xi)\,|\,\vec\xi\in\Omega\wedge
        \xi_1,\ldots,\xi_\nu=0\}
  \eqsep.
\end{equation}
Let $\vec\xi_0$ denote again the parameters under the null hypothesis. Now we
define the hyperplanes $H$ and $H_c$ as tangent planes on $M$ and $M_c$ at the
point $\tilde{\vec x}_0=\tilde{\vec x}(\vec\xi_0)$:
\begin{align}
  H&=\left\{\tilde{\vec x}_0+\vec h\ \vrule\ \vec h\in
     \Span\left( \left.\frac{\partial\tilde{\vec x}(\vec\xi)}
                            {\partial\xi_1}\right|_{\vec\xi=\vec\xi_0},
                 \ldots,
                 \left.\frac{\partial\tilde{\vec x}(\vec\xi)}
                            {\partial\xi_k}\right|_{\vec\xi=\vec\xi_0}\right)
     \right\}
  \eqsep,\nonumber\\
  H_c&=\left\{\tilde{\vec x}_0+\vec h\ \vrule\ \vec h\in
       \Span\left( \left.\frac{\partial\tilde{\vec x}(\vec\xi)}
                              {\partial\xi_{\nu+1}}\right|_{\vec\xi=\vec\xi_0},
                   \ldots,
                   \left.\frac{\partial\tilde{\vec x}(\vec\xi)}
                              {\partial\xi_k}\right|_{\vec\xi=\vec\xi_0}\right)
       \right\}
  \eqsep.\label{eq:nested_hyperplanes}
\end{align}
By construction the function $D(\tilde{\vec x}_0,\vec x)$, considered as a
function of $\vec x$, has a minimum at $\vec x=\tilde{\vec x}_0$ (see
Sec.~\ref{sec:setup}). Consequently, we define the matrix $V^{-1}$ as the
Hessian matrix at that minimum:
\begin{equation}\label{eq:nested_hessian}
  (V^{-1})_{ij} = \left.
  \frac{\partial^2 D(\tilde{\vec x}_0,\vec x)}{\partial x_i\partial x_j}
  \right|_{\vec x=\tilde{\vec x}_0}
  \eqsep.
\end{equation}
As in the derivation of Wilks' theorem, we now perform an affine-linear 
coordinate transformation which maps $\tilde{\vec x}_0$ to zero and transforms
$V^{-1}$ to a unit matrix. To this end, we define
\begin{equation}\label{eq:y}
  y_i\equiv y_i(\vec x)
  = \frac1{\sigma_i}[O(\vec x-\tilde{\vec x}_0)]_i
  \eqsep\text{(no sum over $i$)}\eqsep,
\end{equation}
where $O$ is an orthogonal matrix chosen so that
$OVO^\trans=\diag(\sigma_1^2,\ldots,\sigma_n^2)$ with positive eigenvalues
$\sigma_1^2,\ldots,\sigma_n^2$. Let $H'$ and $H'_c$ denote the images of the
hyperplanes $H$ and $H_c$, respectively, under this coordinate transformation.
Since $H$ and $H_c$ contain $\tilde{\vec x}_0$, $H'$ and $H'_c$ must contain the
origin and are therefore linear subspaces of $\R^n$. Any vector $\vec y$ may
thus be decomposed into three orthogonal components $\vec y_1$, $\vec y_2$ and
$\vec y_3$ with $\vec y_1\in H'_c$ and $\vec y_2\in H$ (see
Fig.~\ref{fig:nested}). If the assumptions of gaussianity and linearity were
satisfied exactly, the theta function in \eqref{eq:pvalueMC} would vanish if and
only if $\|\vec y_2\|^2<S_0$ and we should not waste sample points on this
region. If gaussianity and linearity are only approximations, we should be
more careful and use a sampling density $\rho$ which is small, but non-vanishing
for $\|\vec y_2\|^2<S_0$. A choice which can still be sampled efficiently is
\begin{equation}\label{eq:rho_nested}
  \rho(\vec x) = Je^{-\frac12\|\vec y_1\|^2}e^{-\frac12\|\vec y_3\|^2}
  \begin{cases}
    a\|\vec y_2\|^\alpha &,\eqsep \|\vec y_2\|^2 < S_0\\
    be^{-\frac12\|\vec y_2\|^2} &,\eqsep
      \|\vec y_2\|^2\geq S_0
  \end{cases}
  \eqsep,
\end{equation}
where $J=\prod_{i=1}^n\sigma_i$ is the Jacobian of the coordinate
transformation.  The parameters and $a,b,\alpha\geq0$ may be tuned to improve
the efficiency of the numerical integration (subject, of course, to the
constraint that the PDF $\rho$ is properly normalised).

We see that the sampling density $\rho$ factorises into three terms which only
depend on the components $\vec y_1$, $\vec y_2$ and $\vec y_3$, respectively.
The task of generating points distributed according to $\rho$ thus reduces to
the task of generating components $\vec y_1$, $\vec y_2$ and $\vec y_3$
distributed according to the respective factors.  For $\vec y_1$ and $\vec y_3$
these factors are Gaussian, so generating the components $\vec y_1$ and $\vec
y_3$ is trivial. The generation of the component $\vec y_2$ requires special
care.

\begin{figure}
  \centering\includegraphics{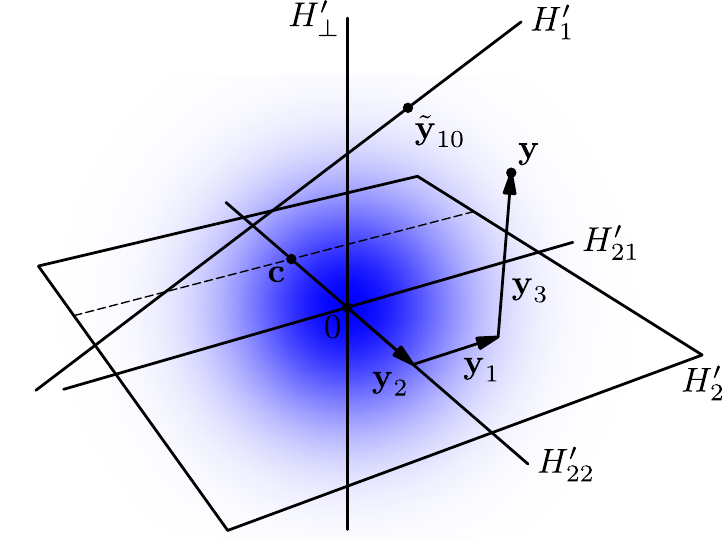}
  \caption{Orthogonal decomposition of a three-dimensional sample space for
    non-nested models. The tangent hyperplane $H'_1$ of model 1 is
    one-dimensional and the tangent hyperplane $H'_2$ of model 2 is
    two-dimensional. The thin dashed line is the projection of $H'_1$ onto
    $H'_2$.  The blue colour indicates the probability density for the toy
    observable vector $\vec y$.}
  \label{fig:nonnested}
\end{figure}

Before we address this problem let us talk about the case of non-nested models.
Assume that we have two theory functions $\tilde{\vec x}_1\equiv\tilde{\vec
  x}_1(\vec\xi)$ and $\tilde{\vec x}_2\equiv\tilde{\vec x}_2(\vec\eta)$ with
parameter spaces $\Omega_1$ and $\Omega_2$, respectively, of arbitrary and
possibly different dimension. Our null hypothesis is that theory 2 is realised
with parameters $\vec\eta_0\in\Omega_2$. We therefore approximate the theory
manifold $M_2$ by its tangent hyperplane $H_2$ at $\tilde{\vec
  x}_{20}\equiv\tilde{\vec x}_2(\vec\eta_0)$. Since $M_2$ is no subset of $M_1$
we have to approximate $M_1$ by its tangent hyperplane at some other parameters
$\vec\xi_0\in\Omega_1$. If $\vec\eta_0$ is the maximum likelihood estimate of
some measured data $\vec x_0$, i.e.\ $\vec\eta_0=\hat{\vec\eta}(\vec x_0)$, an
obvious choice for $\vec\xi_0$ would be the maximum likelihood estimate of $\vec
x_0$ in theory 1, i.e.\ $\vec\xi_0=\hat{\vec\xi}(\vec x_0)$. In any case we
define hyperplanes $H_1$ and $H_2$ analogous to \eqref{eq:nested_hyperplanes}:
\begin{align}
  H_1&=\left\{\tilde{\vec x}_{10}+\vec h\ \vrule\ \vec h\in
       \Span\left( \left.\frac{\partial\tilde{\vec x}_1(\vec\xi)}
                              {\partial\xi_1}\right|_{\vec\xi=\vec\xi_0},
                   \left.\frac{\partial\tilde{\vec x}_1(\vec\xi)}
                              {\partial\xi_2}\right|_{\vec\xi=\vec\xi_0},
                   \ldots\right)\right\}
  \eqsep,\nonumber\\
  H_2&=\left\{\tilde{\vec x}_{20}+\vec h\ \vrule\ \vec h\in
       \Span\left( \left.\frac{\partial\tilde{\vec x}_2(\vec\eta)}
                              {\partial\eta_1}\right|_{\vec\eta=\vec\eta_0},
                   \left.\frac{\partial\tilde{\vec x}_2(\vec\eta)}
                              {\partial\eta_2}\right|_{\vec\eta=\vec\eta_0},
                   \ldots\right)\right\}
  \eqsep,\label{eq:nonnested_hyperplanes}
\end{align}
with $\tilde{\vec x}_{10}=\tilde{\vec x}_1(\vec\xi_0)$ and $\tilde{\vec
  x}_{20}=\tilde{\vec x}_2(\vec\eta_0)$.  We define the matrix $V^{-1}$ as in
\eqref{eq:nested_hessian} and construct coordinates $\vec y\equiv\vec y(\vec x)$
according to \eqref{eq:y}, but with $\tilde{\vec x}_0$ replaced by $\tilde{\vec
  x}_{20}$.  Let $H'_1$ and $H'_2$ be the images of $H_1$ and $H_2$,
respectively, under the coordinate transformation $\vec y$ and let $\tilde{\vec
  y}_{10}=\vec y(\tilde{\vec x}_{10})$. The image of $\tilde{\vec x}_{20}$ under
$\vec y$ is the origin. Since $H_2$ contains $\tilde{\vec x}_{20}$ the
hyperplane $H'_2$ contains the origin and is thus a linear subspace of $\R^n$.
$H'_1$, on the other hand, contains $\tilde{\vec y}_{10}$ but not necessarily
the origin, so it is not a linear subspace. For $n=3$, a two-dimensional $H'_2$
and a one-dimensional $H'_1$, this situation is depicted in
Fig.~\ref{fig:nonnested}.

We see that for non-nested models the boundaries of the region with $S(\vec
x)<S_0$ are curved, even if a linear approximation is used for the theory
manifolds. This makes it harder to construct a sampling distribution which
avoids this region. We shall try it anyway: let $H'_{21}\subset H'_2$ be the
subspace obtained by shifting $H_1$ by $-\tilde{\vec y}_{10}$ (so that it
contains the origin) and projecting it onto $H'_2$. Furthermore, let $H'_{22}$
be the orthogonal complement of $H'_{21}$ in $H_2$. Finally, let $H'_\perp$ be
the orthogonal complement of $H'_2$ in $\R^n$. The projection of $H'_1$ onto
$H'_{22}$ is then a single point $\vec c$, which can be obtained by projecting
$\tilde{\vec y}_{10}$ onto $H'_{22}$. Any vector $\vec y$ can now be written as
the sum of three orthogonal components $\vec y_1$, $\vec y_2$ and $\vec y_3$
with $\vec y_1\in H'_{21}$, $\vec y_2\in H'_{22}$ and $\vec y_3\in H'_\perp$.
(See Fig.~\ref{fig:nonnested}.)  The distance between $\vec y$ and $H'_1$ is
larger than $\|\vec c - \vec y_2\|$ since projecting any vector on a
lower-dimensional subspace reduces its length and the projection of any vector
pointing from $\vec y$ to $H'_1$ onto the subspace $H'_{22}$ is $\vec c - \vec
y_2$.

Now recall the test statistic $S_2$ from \eqref{eq:S2}, which we constructed to
test theory 2 against the ``union'' of theories 1 and 2. In the approximation
where the theory manifolds $M_1$ and $M_2$ are equal to the hyperplanes $H_1$
and $H_2$ the functions $D_1^\text{min}$ and $D_2^\text{min}$ satisfy
\begin{equation}
  D_2^\text{min}(\vec x) = \|\vec y_3\|^2+n\ln(2\pi)
  \eqsep,\eqsep
  D_1^\text{min}(\vec x) \geq \|\vec c-\vec y_2\|^2+n\ln(2\pi)
  \eqsep.
\end{equation}
Thus, $S_2(\vec x)<S_0$ holds if
\begin{equation}\label{eq:nonnested_cond}
  \|\vec y_3\|^2 < S_0 + \|\vec c-\vec y_2\|^2
  \eqsep.
\end{equation}
Note, however, that this condition is sufficient but not necessary.  $S_2(\vec
x)$ must be smaller than $S_0$ in the region defined by
\eqref{eq:nonnested_cond}, but it may also be smaller than $S_0$ outside this
region. Nonetheless, a good choice for the sampling density $\rho$ will
be one which avoids the region defined by \eqref{eq:nonnested_cond}.
Analogous to \eqref{eq:rho_nested}, this density can be constructed as
follows:
\begin{equation}\label{eq:rho_nonnested}
  \rho(\vec x) = Je^{-\frac12\|\vec y_1\|^2}e^{-\frac12\|\vec y_2\|^2}
  \begin{cases}
    a\|\vec y_3\|^\alpha &,\eqsep
      \|\vec y_3\|^2 < S_0 + \|\vec c - \vec y_2\|^2\\
    be^{-\frac12\|\vec y_3\|^2} &,\eqsep
      \|\vec y_3\|^2 \geq S_0 + \|\vec c - \vec y_2\|^2
  \end{cases}
  \eqsep,
\end{equation}
where $J=\prod_{i=1}^n\sigma_i$ is again the Jacobian of the coordinate
transformation $\vec y$. After requiring that $\rho$ is properly normalised,
there are still two free parameters which can be tuned to improve the efficiency
of the numerical integration. As in \eqref{eq:rho_nested}, the density $\rho$
factorises into three terms. The first two are Gaussian and only depend on the
components $\vec y_1$ and $\vec y_2$, respectively. Generating components $\vec
y_1$ and $\vec y_2$ with the correct statistical distribution is therefore
trivial. A new complication in \eqref{eq:rho_nonnested} is that the last factor,
i.e.\ the distribution of $\vec y_3$, now depends on $\vec y_2$.  This simply
means that we have to generate a value for $\vec y_2$ \emph{before} we generate
$\vec y_3$.

The remaining problem is to generate a random vector $\vec z$ of some dimension
$m$ distributed according to the PDF
\begin{equation}
  \rho'(\vec z) = \begin{cases}
    a\|\vec z\|^\alpha &,\eqsep \|\vec z\|^2 < \Delta^2 \\
    be^{-\frac12\|\vec z\|^2} &,\eqsep \|\vec z\|^2 \geq \Delta^2
  \end{cases}
\end{equation}
with some $\Delta > 0$. (In \eqref{eq:rho_nested} we have $\vec z=\vec y_2$,
$m=\nu$ and $\Delta^2=S_0$ while in \eqref{eq:rho_nonnested} we have $\vec
z=\vec y_3$, $m=\dim(H'_\perp)$ and $\Delta^2=S_0 + \|\vec c - \vec y_2\|^2$.)
We first note that the PDF $\rho'$ is rotationally invariant. The length
$r=\|\vec z\|$ of the vector $\vec z$ is then distributed according to a PDF
\begin{equation}
  \tilde\rho'(r) = \frac{2\pi^{m/2}}{\Gamma(m/2)}r^{m-1}\begin{cases}
    ar^\alpha &,\eqsep r^2 < \Delta^2 \\
    be^{-\frac12r^2} &,\eqsep r^2 \geq \Delta^2
  \end{cases}
  \eqsep.
\end{equation}
We may write this as
\begin{equation}
  \tilde\rho'(r)=f\tilde\rho'_<(r) + (1-f)\tilde\rho'_>(r)
\end{equation}
where $f\in[0,1]$ is a free parameter and
\begin{equation}
    \tilde\rho'_<(r)
  = \frac{m+\alpha}{\Delta^{m+\alpha}}\theta(r)\theta(\Delta-r)
  \eqsep,\eqsep
    \tilde\rho'_>(r)
  = \frac{r^{m-1}e^{-\frac12r^2}\theta(r-\Delta)}
         {2^{(m-2)/2}\Gamma(m/2)(1-P_{m/2}(\tfrac12\Delta^2))}
\end{equation}
are PDFs normalised to 1. Here, $P_{m/2}$ is the normalised lower incomplete
Gamma function. Since $\tilde\rho'_<$ and $\tilde\rho'_>$ are normalised, the
parameter $f$ is just the fraction of sample points that will be put in the
``inner region'' with $r<\Delta$. For a given $f$, the corresponding values of
$a$ and $b$ are
\begin{equation}
  a = \frac{f\Gamma(m/2)(m+\alpha)}{2\pi^{m/2}\Delta^{m+\alpha}}
  \eqsep,\eqsep
  b = \frac{1-f}{(2\pi)^{m/2}(1-P_{m/2}(\tfrac12\Delta^2))}
  \eqsep.
\end{equation}
By integrating $\tilde\rho'_<$, $\tilde\rho'_>$ and $\tilde\rho'$ from $0$ to
$r$ we obtain the \emph{cumulative distribution functions} (CDFs)
\begin{gather}
  \CDF_{\tilde\rho'_<}(r) = \frac{r^{m+\alpha}}{\Delta^{m+\alpha}}
  \eqsep,\eqsep
    \CDF_{\tilde\rho'_>}(r)
  = \frac{P_{m/2}(\tfrac12r^2)-P_{m/2}(\tfrac12\Delta^2)}
         {1-P_{m/2}(\tfrac12\Delta^2)}
  \nonumber\\
  \Rightarrow\eqsep
  \CDF_{\tilde\rho'}(r) = \begin{cases}
    f\CDF_{\tilde\rho'_<}(r) &,\eqsep r<\Delta \\
    f + (1-f)\CDF_{\tilde\rho'_>}(r) &,\eqsep r\geq\Delta
  \end{cases}
  \eqsep.
\end{gather}
To generate random variables $r$ distributed according to $\tilde\rho'$ we
need the inverse of $\CDF_{\tilde\rho'}$:
\begin{gather}
  \CDF^{-1}_{\tilde\rho'_<}(q) = \Delta\,q^{1/(m+\alpha)}
  \eqsep,\eqsep
    \CDF^{-1}_{\tilde\rho'_>}(q)
  = \sqrt{2P^{-1}_{m/2}\bigl(q + (1-q)P_{m/2}(\tfrac12\Delta^2)
     \bigr)}
  \nonumber\\
  \Rightarrow\eqsep
  \CDF^{-1}_{\tilde\rho'}(q) = \begin{cases}
    \CDF^{-1}_{\tilde\rho'_<}(\tfrac qf) &,\eqsep q<f \\
    \CDF^{-1}_{\tilde\rho'_>}(\tfrac{q-f}{1-f}) &,\eqsep q\geq f
  \end{cases}
  \eqsep,
\end{gather}
where $P^{-1}_{m/2}$ is the inverse of the normalised lower incomplete Gamma
function. Random vectors $\vec z$ distributed according to $\rho$ may now be
generated in the following way: First generate a vector $\vec z'$ according to
a $m$-dimensional normal distribution. Then pick a uniformly distributed random
variable $q\in[0,1]$ and set $r=\CDF^{-1}_{\tilde\rho'}(q)$. The variable $r$ is
then distributed according to $\tilde\rho'$. The vector $\vec z$ with the
correct random distribution is $\vec z=(r/\|\vec z'\|)\vec z'$.

\section{Introducing \myFitter}
\label{sec:myFitter}

The ideas for the numerical computation of $p$-values outlined in the last
section have been implemented in a publicly available code called \myFitter.
The source code is available at Hepforge \cite{myFitter}. Detailed documentation
is included in the source distribution. Here I just want to provide a brief
description of the user interface and discuss some details of the
implementation.

{\myFitter} is a C++ class library and makes extensive use of inheritance and
polymorphism to separate the tasks of fitting a model to experimental data
and computing $p$-values from the tasks of implementing the observables (as
functions of the model's parameters) or the input function $D$ (as a
function of the observables). The main classes the user will have to
deal with are:
\begin{description}
\item[\code{Model}] This is the base class for all models implemented by the
  user. It essentially represents the theory function $\tilde{\vec x}$ from
  earlier sections, i.e.\ the map from the model's parameter space to the space
  of observables. The base class provides functionality for storing ``current''
  values of parameters, observables and derivatives of observables with respect
  to the parameters, setting ranges in which parameters are allowed to float or
  fixing them (so that they do not float at all).  It can also randomly sample
  the parameter space and build up a dictionary of parameter values and the
  corresponding observable values. This dictionary can be used to find good
  starting points for numerical minimisations of the input function. To
  implement their own model, the user has to subclass \code{Model} and overload
  the method \code{calc()} which computes the observables based on the current
  values of the parameters. They may also overload the method
  \code{calc\_deriv()}, which calculates the derivatives of all observables with
  respect to all parameters.  The default implementation uses simple numerical
  differentiation.
\item[\code{InputComponent}] This is the base class for objects that represent
  terms in the input function $D$ (see Sec.~\ref{sec:setup}). Each input
  component represents the contribution from one or more observables $x_i$ to
  the input function. To calculate the value of the input function, the
  contributions of all \code{InputComponent} objects are added up. This is done
  by another class, \code{InputFunction}, which acts as a container for
  \code{InputComponent} objects. Derived classes of \code{InputComponent} must
  overload the method \code{calc($\tilde{\vec x},\vec x$)}, which takes two
  vectors as arguments (the first being the ``predicted'' values of the
  observables and the second being the ``measured'' ones) and returns the
  contribution of the term to the input function.  Additionally, the methods
  \code{calc\_deriv()} and \code{get\_hessian()} must be implemented, which
  calculate the derivatives with respect to the $\tilde x_i$ and the Hessian
  matrix for the minimum at $\vec x=\tilde{\vec x}$.  Ready-to-use
  implementations for the most common input components are also available. These
  classes are: \code{GaussianIC} (for single observables with a Gaussian and
  possibly systematic errors), \code{AsymmetricGaussianIC} (for single
  observables with asymmetric Gaussian error bars and possibly systematic
  errors) and \code{CorrelatedGaussianIC} (for several observables with Gaussian
  errors and a correlation matrix).
\item[\code{Fitter}] Objects of this type are responsible for fitting the
  parameters of models (represented by \code{Model} objects) to experimental
  data (represented by an \code{InputFunction} object) and for computing
  $p$-values by numerical integration. Each \code{Fitter} object contains an
  \code{InputFunction} object which is accessible through the
  \code{input\_function()} method and must be ``filled'' with
  \code{InputComponent} objects before any fits can be done. Once the input
  function is initialised, fits can be performed with the \code{local\_fit()}
  and \code{global\_fit()} methods. As arguments, these methods take the
  \code{Model} object to be fitted and (optionally) a vector of central values
  for the observables. If no central values are given, the defaults from the
  \code{input\_function()} are used.  The difference between these methods is
  that \code{local\_fit()} uses the current values of the model parameters as
  starting point for the minimisation of the input function, while
  \code{global\_fit()} uses the dictionary created by a previous call to the
  model's \code{scan()} method. The $p$-values for likelihood ratio tests of
  nested and non-nested models can be calculated with the methods
  \code{calc\_nested\_lrt\_pvalue()} and \code{calc\_lrt\_pvalue()},
  respectively.  As arguments, these two methods take the models to be
  compared. Note that, for \code{calc\_nested\_lrt\_pvale()} to work, the second
  model must be a restricted version of the first, i.e.\ a copy of the first
  object with some additional parameters fixed. In addition to these methods,
  the \code{Fitter} class contains numerous options and flags that control the
  accuracy and various other aspects of the minimisation and integration
  routines. These options are described in the package documentation. Most
  notably, the $p$-value integrations can be \emph{parallelised without
    additional programming efforts} by the user.
\end{description}

Both, for the case of nested and non-nested models, the efficiency of the
integration can be improved further by \emph{adaptive} integration techniques,
where the shape of the sampling density $\rho$ is tuned \emph{during} the actual
integration. For the adaptation, the implementation in {\myFitter} uses the
OmniComp/Dvegas package \cite{DVEGAS} by Nikolas Kauer, which implements the
VEGAS algorithm \cite{VEGAS} and was developed in the context of
\cite{Kauer:2001sp,Kauer:2002sn}. Thanks to OmniComp, parallelised integration
is fully supported.

To maximise the likelihood function, {\myFitter} uses a custom implementation of
the BFGS method for numerical optimisation \cite{Broyden, Fletcher, Goldfarb,
  Shanno}. The optimisation terminates successfully when the length of the
gradient of the likelihood function drops below a certain value configurable by
the user. Other optimization algorithms can be implemented by subclassing the
\code{Minimizer} class and assigning an instance of this class to the
\code{Fitter} object via the \code{Fitter::minimizer()} method. The problem of
minimising a function of bounded parameters (i.e.\ of parameters that have an
upper or lower limit) is solved in the usual way by smoothly and invertably
mapping the real axis $\R$ to the allowed range of the parameter.  Internally,
{\myFitter} does this with the function
\begin{equation}
  g:(-\infty,\infty)\to(0,\infty),\ x\mapsto f(x) = \frac12(x + \sqrt{x^2+1})
  \eqsep.
\end{equation}

\section{Performance Tests}
\label{sec:performance}

The performance of the {\myFitter} method for the numerical integration of
Eq.~\ref{eq:pvalue} was compared with more generic methods in three tests using
simple toy models. All alternative methods use the coordinate transformation
\eqref{eq:y} to transform the PDF $f$ to a normal distribution.  The simplest
method, referred to as \emph{no adaptation} in the following, just uses
importance sampling with a normal distribution as sampling density.  The other
two methods map the integration volume (i.e.\ the $\R^n$) to the unit hypercube
$[0,1]^n$ by using
\begin{equation}
  t_i = \frac12\Erf\frac{y_i}{\sqrt 2} + \frac12
  \eqsep,\eqsep i=1,\ldots,n
\end{equation}
as integration variables and use the VEGAS algorithm \cite{VEGAS} to perform the
integration over the variables $t_i$. The VEGAS algorithm is most efficient when
the features of the integrand are aligned with the coordinate axes. In one
variant, called \emph{aligned VEGAS} in the following, we perform a rotation
which aligns the tangent hyperplanes of the theory manifolds with the
coordinate axes before mapping to the unit cube. This usually leads to an
integrand whose features are aligned with the coordinate axes. In a second
variant, which we call \emph{misaligned VEGAS}, we choose the rotation so that
the theory manifolds are \emph{not} aligned with the coordinate axes. The
misaligned VEGAS method is the best possible method when no information about
the theory manifolds can be used.

In the first test we study the performance of the four integration methods in
the context of a model with a curved theory manifold.  To this end we consider a
model with seven observables $x_1,\ldots,x_7$ and four parameters
$\xi_1,\ldots,\xi_4$. The theory function $\tilde{\vec x}$ is given by
\begin{equation}\label{eq:test_theory}
  \tilde{\vec x}(\vec\xi)
  =(\xi_1,\xi_2,\xi_3,\xi_4,0,0,-(\xi_1^2+\xi_2^2+\xi_3^2+\xi_4^2)\lambda)
  \eqsep.
\end{equation}
where $\lambda$ is a fixed number which controls the curvature of the theory
manifold. As input function we use the expression \eqref{eq:gaussian} for
Gaussian errors with a unit covariance matrix:
\begin{equation}\label{eq:test_input}
  D(\tilde{\vec x},\vec x) = \sum_{i=1}^7(\tilde x_i-x_i)^2 + 7\ln(2\pi)
  \eqsep.
\end{equation}
The constrained version of this model is defined by fixing $\xi_2$ to zero and
$\xi_1$ to some other value $r$. The test statistic $S$ is then defined
according to \eqref{eq:S}. We take $\vec x_0=(0,0,0,0,1,1,1)$ as the actually
measured data and perform the test with $S_0=S(\vec x_0)$. Different choices for
$r$ lead to different values of $S_0$ and thus to different $p$-values.

\begin{table}
\centering
\begin{tabular}{CCCrrrr}
  \toprule
  & & & 
  & \multicolumn{1}{c}{aligned}
  & \multicolumn{1}{c}{misaligned}
  & \multicolumn{1}{c}{no}
  \\
  \lambda & \multicolumn{1}{c}{$p$-value} & \multicolumn{1}{c}{$p$-value}
  & \multicolumn{1}{c}{{\myFitter}}
  & \multicolumn{1}{c}{VEGAS}
  & \multicolumn{1}{c}{VEGAS}
  & \multicolumn{1}{c}{adaptation}
  \\
  & & \multicolumn{1}{c}{(Wilks)} 
  & \multicolumn{1}{c}{$[10^3]$}
  & \multicolumn{1}{c}{$[10^3]$}
  & \multicolumn{1}{c}{$[10^3]$}
  & \multicolumn{1}{c}{$[10^3]$}
  \\\midrule[\heavyrulewidth]
  0.1 & 5.1\cdot 10^{-2} & 5.6\cdot 10^{-2} &  40 &   60 &   90 &  200 \\
      & 2.9\cdot 10^{-3} & 3.3\cdot 10^{-3} & 150 &  240 &  360 &  --\ \\
      & 6.1\cdot 10^{-5} & 8.1\cdot 10^{-5} & 210 & 2100 & 3000 &  --\ \\
      & 5.4\cdot 10^{-7} & 7.3\cdot 10^{-7} & 270 &  --\ &  --\ &  --\ \\
  \midrule
  1.0 & 5.2\cdot 10^{-2} & 8.8\cdot 10^{-2} &  50 &   60 &   70 &  200 \\
      & 3.0\cdot 10^{-3} & 5.8\cdot 10^{-3} & 180 &  210 &  210 &  --\ \\
      & 8.2\cdot 10^{-5} &20.1\cdot 10^{-5} & 700 & 2100 & 1800 &  --\ \\
      & 8.1\cdot 10^{-7} &29.7\cdot 10^{-7} &6000 &  --\ &  --\ &  --\ \\ 
  \bottomrule
\end{tabular}
\caption{Results of the test with a curved theory manifold. The curvature is
  controlled by $\lambda$ (see Eq.~\ref{eq:test_theory}). The parameter $\xi_1$
  was fixed to different values in the constrained model, leading to the
  $p$-values shown in the second column (which roughly correspond to $2$, $3$,
  $4$ and $5$ standard deviations). The $p$-values obtained by applying Wilks
  theorem is shown in the third column.  The numbers in the last four columns
  are the number of integrand evaluations needed by the four different
  integration methods to compute the $p$-value with a relative accuracy of
  $1\%$. In the empty cells, the integration was aborted after a number of
  evaluations which was a factor of 10 larger than the evaluations needed by
  slowest of the other methods.}
\label{tab:curvature}
\end{table}
For two values of $\lambda$, the value of $r$ was varied to obtain $p$-values
roughly corresponding to $2$, $3$, $4$ and $5$ standard deviations. The
$p$-value was then computed numerically with {\myFitter} and the three
alternative methods to a relative precision of $1\%$. The number of integrand
evaluations needed by each method are summarised in
Tab.~\ref{tab:curvature}. The $p$-values obtained by applying Wilks theorem are
also shown. We see that adaptive methods always lead to a significant speedup
and that the {\myFitter} method performs best in all cases. For three standard
deviations or less the two VEGAS methods still compete rather well with
{\myFitter}. At four standard deviations {\myFitter} is faster than the VEGAS
methods by a factor of 3 for large curvature and a factor of 10 for small
curvature. At five standard deviations only {\myFitter} is able to compute the
$p$-value with a reasonable number of evaluations.  The main reason for the poor
performance of the VEGAS methods at small $p$-values is the fact that they
require a large number of initial evaluations to find \emph{any} points in the
integration region which give a nonzero contribution to the integrand. The
{\myFitter} method converges faster because it ``knows'', to a certain
approximation, where the integrand is nonzero.

The second test compares the performance of the four integration methods in the
case of models with bounded parameters. To this end, we use the theory function
\eqref{eq:test_theory} with $\lambda=0$ and the input function
\eqref{eq:test_input}. In the constrained version of the model, the parameters
$\xi_1$ and $\xi_2$ are still fixed to $r$ and $0$, respectively.  We assume
again that $\vec x_0=(0,0,0,0,1,1,1)$ is the actually measured data, perform the
fit with $S_0=S(\vec x_0)$ and vary $r$ to change the $p$-value.  However, in
the full model the parameter $\xi_2$ is now restricted to the interval $[-0.25,
  0.25]$. Thus, the ``effective'' number of degrees of freedom of the LRT is
somewhere between one and two. Consequently, we expect the $p$-value to lie
somewhere between the results obtained from Wilks theorem with one and two
degrees of freedom.

\begin{table}
\centering
\begin{tabular}{CCrrrr}
  \toprule
  & & 
  & \multicolumn{1}{c}{aligned}
  & \multicolumn{1}{c}{misaligned}
  & \multicolumn{1}{c}{no}
  \\
  \multicolumn{1}{c}{$p$-value} & \multicolumn{1}{c}{$p$-value}
  & \multicolumn{1}{c}{{\myFitter}}
  & \multicolumn{1}{c}{VEGAS}
  & \multicolumn{1}{c}{VEGAS}
  & \multicolumn{1}{c}{adaptation}
  \\
  & \multicolumn{1}{c}{(Wilks)} 
  & \multicolumn{1}{c}{$[10^3]$}
  & \multicolumn{1}{c}{$[10^3]$}
  & \multicolumn{1}{c}{$[10^3]$}
  & \multicolumn{1}{c}{$[10^3]$}
  \\\midrule[\heavyrulewidth]
  4.4\cdot 10^{-2} & 11.0\cdot 10^{-2} &  30 &   30 &  110 &  240 \\
  2.8\cdot 10^{-3} &  9.5\cdot 10^{-3} &  30 &   50 &  460 & 4100 \\
  6.3\cdot 10^{-5} & 27.4\cdot 10^{-5} &  30 & 1500 & 5100 &   -- \\
  5.4\cdot 10^{-7} & 29.0\cdot 10^{-7} &  40 &   -- &   -- &   -- \\
  \bottomrule
\end{tabular}
\caption{Results of test with bounded parameters (see text). The parameter
  $\xi_1$ was fixed to different values in the constrained model, leading to the
  $p$-values shown in the first column (which roughly correspond to $2$, $3$,
  $4$ and $5$ standard deviations). The $p$-values obtained by applying Wilks
  theorem (with two degrees of freedom) is shown in the second column.  The
  numbers in the last four columns are the number of integrand evaluations
  needed by the four different integration methods to compute the $p$-value with
  a relative accuracy of $1\%$. In the empty cells, the integration was aborted
  after a number of evaluations which was a factor of 10 larger than the
  evaluations needed by slowest of the other methods.}
\label{tab:boundary}
\end{table}
Different $p$-values roughly corresponding to $2$, $3$, $4$ and $5$ standard
deviations were again computed with the four integration methods to a relative
precision of $1\%$. The required number of integrand evaluations are shown in
Tab.~\ref{tab:boundary}. Again the {\myFitter} method performs best in all
cases. The `misaligned VEGAS' and `no adaptation' methods are significantly
slower even for large $p$-values. The performance of the `aligned VEGAS' method
is comparable at first, but drops significantly between $3$ and $4$ standard
deviations. The reason is the same as in the previous test: for small $p$-values
VEGAS needs a large number of initial evaluations in order to find enough points
with a nonzero integrand value. Again, only {\myFitter} is capable of computing
$p$-values at the $5\sigma$ level.

The final test is concerned with the case of non-nested models. The input
function (for both models) is again given by \eqref{eq:test_input}.  The first
model has two parameters $\xi_1$, $\xi_2$ and the theory function $\tilde{\vec
  x}_1$ is defined by
\begin{equation}
  \tilde{\vec x}_1(\vec\xi)=(\xi_1,\xi_2,0,0,\xi_1,\xi_2,0)
  \eqsep.
\end{equation}
The theory function $\tilde{\vec x}_2$ of the second model is the same as
$\tilde{\vec x}$ from \eqref{eq:test_theory} with $\lambda=0$. Obviously,
neither theory manifold contains the other as a subset, so this is an example
of non-nested models. We assume the actually measured data to be
\begin{equation}
  \vec x_0 =(r,r,1,0,r,r,1)
\end{equation}
with $r>0$. For sufficiently large values of $r$ the maximum likelihood value of
model 1 at $\vec x_0$ is larger than that of model 2 and we have the interesting
situation that the model with less parameters fits the measured data better than
the model with more parameters. In this situation Wilks' theorem is clearly not
applicable, so we will concentrate on this case. We perform a LRT which compares
model 2 with the `union' of models 1 and 2, using the test statistic $S_2$ from
\eqref{eq:S2} and $S_0=S_2(\vec x_0)$.

\begin{table}
\centering
\begin{tabular}{CCrrrr}
  \toprule
  & & 
  & \multicolumn{1}{c}{aligned}
  & \multicolumn{1}{c}{misaligned}
  & \multicolumn{1}{c}{no}
  \\
  \multicolumn{1}{c}{$p$-value} & \multicolumn{1}{c}{$p$-value}
  & \multicolumn{1}{c}{{\myFitter}}
  & \multicolumn{1}{c}{VEGAS}
  & \multicolumn{1}{c}{VEGAS}
  & \multicolumn{1}{c}{adaptation}
  \\
  & \multicolumn{1}{c}{(Wilks)} 
  & \multicolumn{1}{c}{$[10^3]$}
  & \multicolumn{1}{c}{$[10^3]$}
  & \multicolumn{1}{c}{$[10^3]$}
  & \multicolumn{1}{c}{$[10^3]$}
  \\\midrule[\heavyrulewidth]
  5.0\cdot 10^{-2} & 50.3\cdot 10^{-2} &  50 &   70 &  100 &  200 \\
  2.7\cdot 10^{-3} & 44.6\cdot 10^{-3} &  60 &  210 &  270 & 4000 \\
  6.7\cdot 10^{-5} &147.0\cdot 10^{-5} &  70 &  330 &  540 &   -- \\
  5.7\cdot 10^{-7} &157.3\cdot 10^{-7} &  80 &   -- &   -- &   -- \\
  \bottomrule
\end{tabular}
\caption{Results of test for non-nested models. The $p$-values in the first
  column were obtained by numerical integration with different values for $r$
  (see text). The $p$-values in the second column were computed by applying
  Wilks' theorem with two degrees of freedom. The numbers in the last four
  columns are the number of integrand evaluations needed by the four different
  integration methods to compute the $p$-value with a relative accuracy of
  $1\%$. In the empty cells, the integration was aborted after a number of
  evaluations which was a factor of 10 larger than the evaluations needed by
  slowest of the other methods.}
\label{tab:nonnested}
\end{table}
As before, several LRTs were performed with different values of $r$ leading to
$p$-values roughly corresponding to $2$, $3$, $4$ and $5$ standard deviations.
The $p$-values were again computed with the four integration methods to a
relative precision of $1\%$, and the required number of integrand evaluations
are shown in Tab.~\ref{tab:nonnested}. The $p$-values obtained by applying
Wilks' theorem with two degrees of freedom are also shown for illustration.  We
see that Wilks' theorem is clearly not applicable here. As in the previous
tests, the {\myFitter} method is consistently faster than the other methods and
significantly faster for small $p$-values.

\section{Conclusions}
\label{sec:concl}

Likelihood ratio tests are a popular tool in global analyses of models in
particle physics. For a correct statistical interpretation of the data, reliable
methods for the computation of $p$-values in likelihood ratio tests are needed.
There are many realistic situations where Wilks' theorem does not apply and the
distribution of the test statistic is not known analytically. These include
likelihood ratio tests of non-nested models or models with parameters that are
only allowed to float in a finite range. Real-world examples of the former case
are the global analyses \cite{Eberhardt:2012ck, Eberhardt:2012gv} of the
Standard Model with a fourth generation of fermions where the models being
compared are not nested due to the non-decoupling nature of the additional
fermions. The latter case includes models where systematic errors are treated
within the $R$Fit scheme. In these situations one has to resort to numerical
methods. Monte Carlo integration can be used to compute $p$-values numerically,
but the integration usually becomes very inefficient for small $p$-values.

In this paper I presented an efficient approach to the numerical computation of
$p$-values which is based on importance sampling and applies to a broad class of
statistical models. In global analyses in particle physics, the predictions of a
theoretical model can be described by a manifold in the space of
observables. The PDF of the statistical model is then obtained by ``smearing
out'' the theory manifold in a way determined by the experimental
uncertainties. The proposed methods use geometric information about the theory
manifolds to construct suitable sampling densities for the Monte Carlo
integration and substantially improve the performance of the numerical
integration for small $p$-values. These methods are implemented in a publicly
available C++ framework for likelihood ratio tests called {\myFitter}.

\section*{Acknowledgements}

I would like to thank J\'er\^ome Charles for fruitful discussions about
likelihood ratio tests for non-nested models. I also thank Otto Eberhardt for
checking fit results with CKMfitter and Ulrich Nierste for thorough proof
reading.

\bibliography{pvalues}
\end{document}